# Flux Jumps Driven by a Pulsed Magnetic Field

J. Vanacken, L. Trappeniers, K. Rosseel, I.N. Goncharov*,
V.V. Moshchalkov and Y. Bruynseraede

*Laboratorium voor Vaste-Stoffysica en Magnetisme, K.U.Leuven, Celestijnenlaan 200 D, B-3001 Leuven, Belgium*
**Laboratory of High Energy, Joint Institute for Nuclear Research, 141980 Dubna, Russia*

**Abstract**

The understanding of flux jumps in the high temperature superconductors is of importance since the occurrence of these jumps may limit the perspectives of the practical use of these materials. In this work we present the experimental study of the role of heavy ion irradiation in stabilizing the HTSC against flux jumps by comparing un-irradiated and $7.5 \bullet 10^{10}$ Kr-ion/cm$^2$ irradiated $(Y_xTm_{1-x})Ba_2Cu_3O_7$ single crystals. Using pulsed field magnetization measurements, we have applied a broad range of field sweep rates from 0.1T/s up to 1800 T/s to investigate the behavior of the flux jumps. The observed flux jumps, which may be attributed to thermal instabilities, are incomplete and have different amplitudes. The flux jumps strongly depend on the magnetic field, on the magneto-thermal history of the sample, on the magnetic field sweep rate, on the critical current density $j_c$, on the temperature and on the thermal contact with the bath in which the sample is immersed.

## 1. Introduction

In studying the influence of columnar defects, created by heavy-ion irradiation, on the magnetic properties of $(Y_xTm_{1-x})Ba_2Cu_3O_7$ superconducting single crystals by pulsed field magnetization measurements, we observed the presence of strong magnetic flux jumps. The use of the pulsed field experimental method is motivated by the large characteristic critical fields present in high temperature superconductors, certainly below a reduced temperature $t = T/T_c = 0.8$. In order to reach magnetic fields up to 60 Tesla, large sweep rates are inherent in this experimental method. These magnetic field sweep rates may reach values as high as 30 kT/s, and give rise to magneto-thermal instabilities, even in relatively small samples (V ~ 0.1 mm$^3$). During a magnetic field sweep, a small perturbation may cause power dissipation ʒ$W_1$ (due to thermally activated (giant) flux creep, flux flow, geometric current distribution inside the sample or due to the specific configuration of the pinning centers, etc…) that leads to an increase of the temperature $\Delta T_1$. This in turn influences the superconductor, and new dissipation $\Delta W_2$ occurs thus resulting in a new the raise of temperature $\Delta T_2$. If $\Delta T_2 < \Delta T_1$ then the next instability $\Delta T_3$ will be smaller than $\Delta T_2$, and the superconductor will recover eventually its original state. If on the other hand the induced increase of temperature $\Delta T_2$ is larger than the original fluctuation $\Delta T_1$, then an instability will be developed and the magnetic energy will be converted into heat. This gives rise to a flux movement in the form of an avalanche. In classical superconductors, flux jumps can easily result in the complete destruction of the superconducting state [3], but in the high $T_c$ materials, very large temperature increases are needed ($\Delta T \sim 100K$) before this effect takes place, and therefore such observations have not been made so far. On the other hand, the thermal instabilities are much higher in high $T_c$ compounds, which makes the observation of partial flux jumps in these materials very common.

## 2. Experimental method

The K.U. Leuven pulsed field facility [1] allows to perform magnetization measurements in fields up to 60 tesla, at temperatures down to 350 mK. The sensitivity of the homemade susceptometer is better than $\Delta m=10^{-3}$ emu at fields below 20 tesla and $10^{-2}$ emu at higher fields [1]. In this work we study an (i) un-irradiated $(Y_xTm_{1-x})Ba_2Cu_3O_7$ single crystal with x=0.14, and an irradiated sample of the same stoichiometry with an irradiation dose of $7.5 \bullet 10^{10}$ Kr-ion/cm$^2$. The single crystals are prepared using the standard flux growth method. Figure 1 shows the comparison of the magnetic critical current densities of both samples at temperature T=50K. At the lower fields (B<4 T) the irradiated sample shows a larger critical current density than the reference un-irradiated sample. This observation is also valid for the lower temperatures.

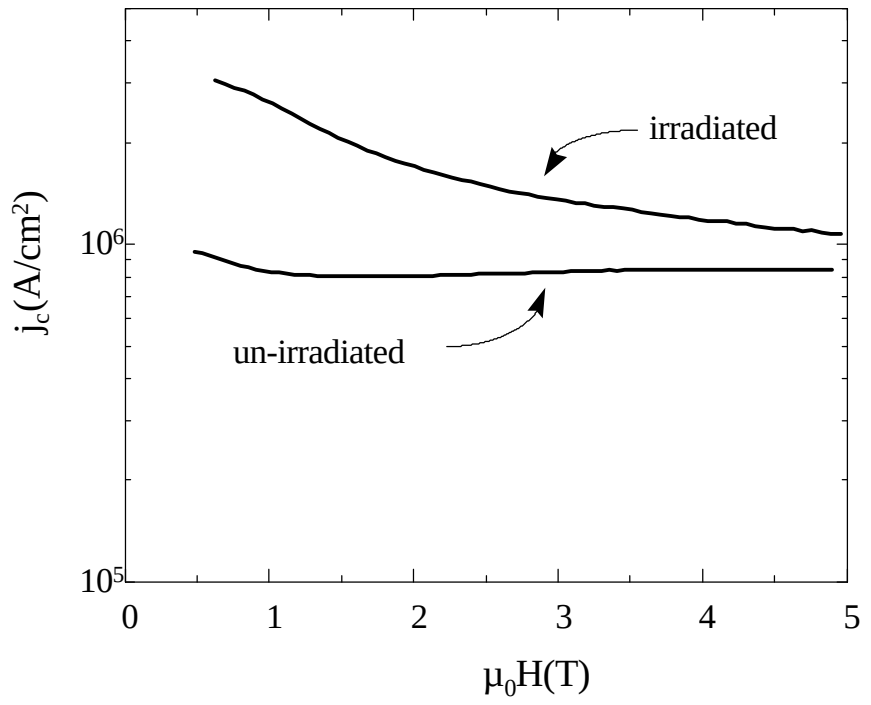

Fig. 1. Comparison of the magnetic critical current density at temperature T=50K of the un-irradiated and the $7.5 \bullet 10^{10}$ Kr-ion/cm$^2$ irradiated sample.

## 3. Experimental results and discussion

### *3.1. Un-irradiated sample*

Figures 2a and 2b show the magnetization versus field for the un-irradiated reference sample at T=4.2K (contact gas). The thin line curve in figure 2a corresponds to the field sweep rate versus field at which the experiment was performed. The arrows indicate the sweep direction. During this experiment the field sweep rate varies from ~1800 T/s down to ~ 0.1 T/s. Initially, the magnetization loop opens normally, and we can estimate the full flux penetration field $\mu_0H^* \sim 1$ T. When the magnitude of the field is decreased sharp jumps appear around H*.

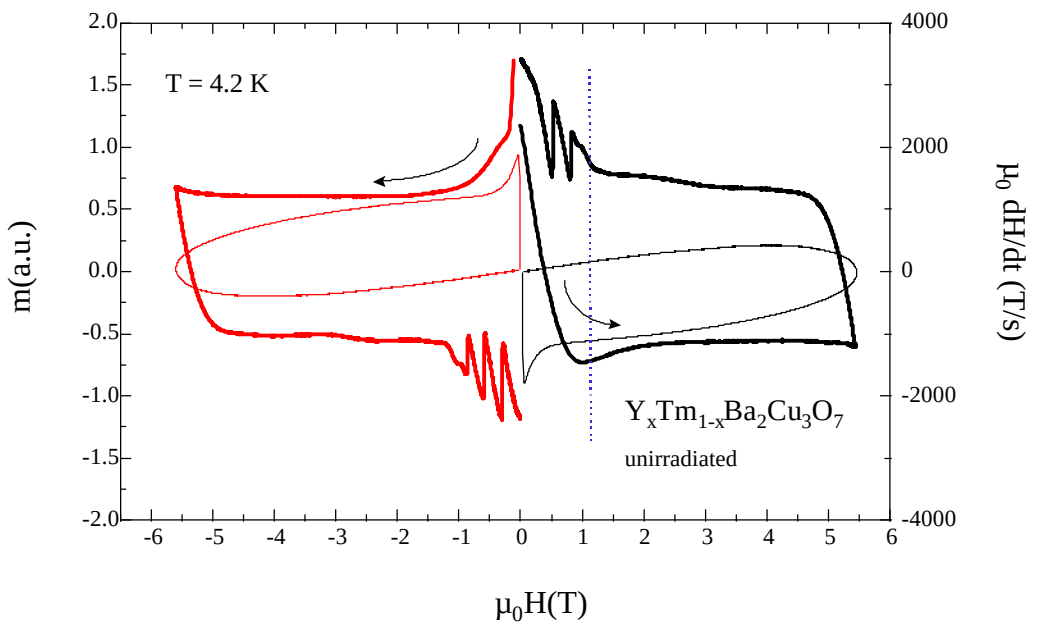

Fig. 2a. The bold line shows the magnetization versus magnetic field for $(Y_xTm_{1-x})Ba_2Cu_3O_7$ un-irradiated single crystals at temperature T=4.2K, whereas the thin line shows the magnetic field sweep rate versus the magnetic field

Details of these experiments are given in figure 2b which illustrates the time dependence of such a flux jump. Only a few, seemingly periodical jumps are observed, for both the positive as the negative field polarities. These sharp flux jumps occur at the end of the field sweep (in lowering the magnitude of the magnetic field) for all temperatures below T=30K. This effect does not demonstrate any obvious field sweep history dependence. In these experiments, a fixed field interval between the flux jumps of $\mu_0H=0.3$ T is observed. Although such a periodicity is not universal, it has been described by Swartz [2] using the following expression [which was derived from the Bean critical state model, i.e. $j(T,H) = j_c(T)$. By taking into account the empirical relation $j_c(T) \sim j_{c0} \exp(T/T_0)$] we further simplified Eq.(1)]:

$$\Delta B = \dot{B}\,\Delta t \leq \sqrt{\frac{m_0\ C_v\ j_c}{\left| \partial\ j_c / \partial T \right|}} = B_j \approx \sqrt{m_0\ C_v\ T_0} \qquad (1)$$

where $\Delta B$ is the field difference between the applied field and the field at the center of the sample, jc is the critical current density and $C_v$ the specific heat. Eq. (1) is only valid if the flux front penetrates the sample sufficiently fast, i.e. when the vortex front speed $v_f$ obeys:

$$v_f \geq \left( \frac{\Delta B}{m_0\ j_c} \right) \frac{1}{\Delta t} \approx \frac{\dot{B}}{m_0\ j_c} \qquad (2)$$

Using pulsed field conditions $\dot{B}$=1 kT/s and $j_c$=$10^8$A/m$^2$, we find $v_f$ ~ 1 km/s, a velocity which is realistic and which indicates that Eq.(2) can be satisfied. As it follows from Eq. (1) a flux jump occurs each time a field difference $\Delta B > B_j$ is applied. In this way the jumps are expected to be equidistant ~ $B_j$. By taking experimental values $T_0$~30K and $C_v$ ~ 3000 WsK$^{-1}$m$^{-3}$ we find $B_j$ ~ 0.3 T as observed on the un-irradiated sample (see figure 2b).

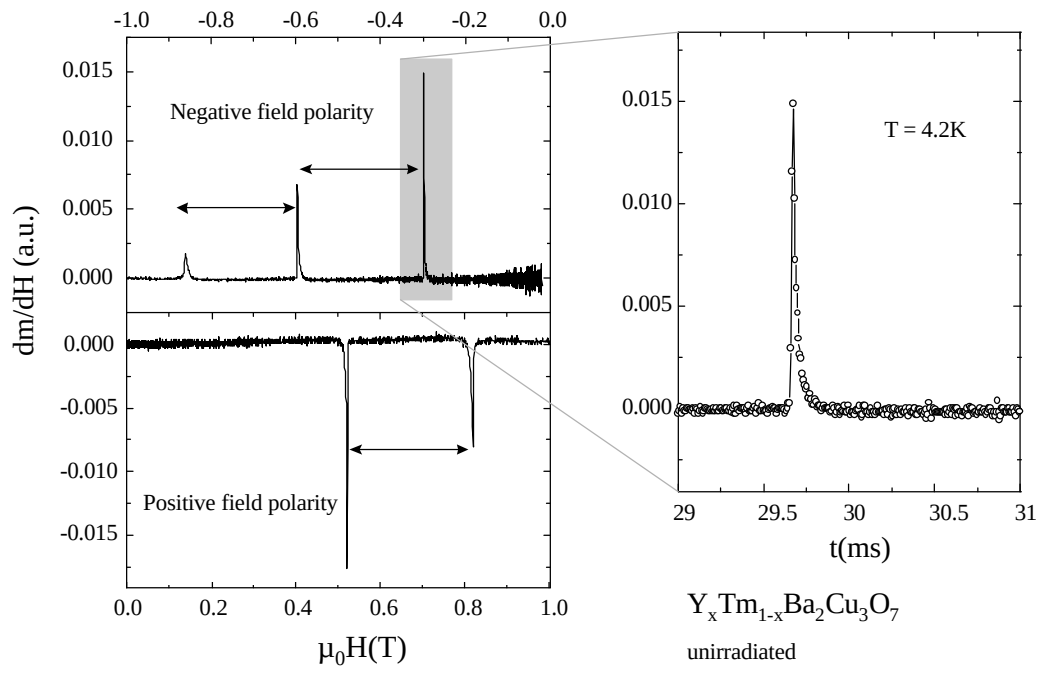

Fig. 2b. The dynamic susceptibility at the end of the field sweep, for both the negative and positive field sweeps (the graphs at the left). The right graph shows the time dependence of the dynamic susceptibility.

The graph at the right of figure 2b shows the dynamic susceptibility $\chi$=dm/dH versus time. The $\chi(H)$ amplitude diverges when the instability occurs. The rise time of the instability is $t_{rise}$ ~ 10 μsec, whereas the lowering time $t_{lower}$ ~ 100 μsec. A typical time constant $\tau_B$~ 50 μsec for such a jump can be defined in this way. The observation of flux movement within such a small time interval is novel and it has become possible due to the use of fast data acquisition systems (~1 MHz sampling).

### 3.2. *Irradiated sample*

The magnetization versus magnetic field for the irradiated $(Y_xTm_{1-x})Ba_2Cu_3O_7$ single crystal at T=4.2K is given in fig. 3. In this graph, curves (1) and (2) correspond to almost identical experiments, the difference being that curve (1) was measured after switching the polarity of the magnetic field, whereas curve (2) is a second measurement performed at the same field polarity. As such, the flux jump effect is clearly dependent upon the field sweep sequence. The occurrence of the jumps is completely reproducible, although the details of the flux jumps (i.e. the fields at which they appear, the magnitude of the jump) are different. Note that the flux jumps are seen at the beginning of the field sweep, and if they appear, the width of the magnetization loop becomes smaller, which is a clear indication that the jumps have raised the overall temperature of the sample (estimated from the $J_c(T)$ dependence to be $\Delta T$ ~ 5K). The insert of this figure shows the dynamic susceptibility $\chi$=dm/dH, which diverges when the flux jumps occur. The rising and lowering time constant of these flux jumps are both around 20 μsec, and the jumps are more symmetric than in the case of the un-irradiated sample. Equidistant flux jumps are not observed (see the insert in Fig.3).

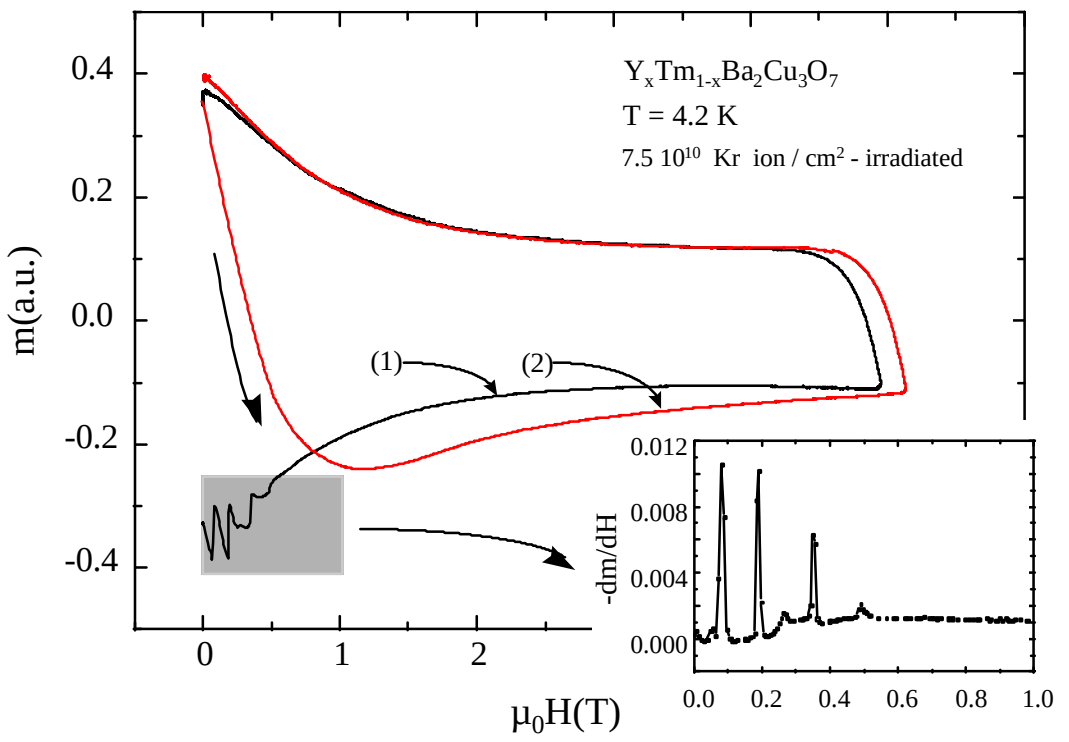

Fig. 3. Magnetization versus magnetic field for a 7.5 • $10^{10}$ ion/cm$^2$ Kr-irradiated ($Y_xTm_{l\ x})Ba_2Cu_30_7$ single crystal at temperature T=4.2K. Curve (1) was measured after switching the polarity of the magnetic field, whereas curve (2) is a second measurement performed at the same field polarity.

Figure 4 shows the magnetization data at T=1.7K. We clearly see a large difference in overall behavior at temperatures T=4.2K and T=1.7K. In the latter experiment the sample was emerged in superfluid He. At T=1.7K the sample shows flux jumps both in the field increasing and decreasing branches. The details of the jumps are illustrated by the graph at the right side, which shows dm/dH versus the magnetic field, for the increasing (upper graph) and decreasing (lower graph) branch. The coupling of the sample to the superfluid helium bath is apparently the cause of this difference between the m(H) behavior at T=4.2K and T=1.7K. If the heating power released per unit volume can be removed by the heat conductivity, then Eq (1) must be changed [2-4] to:

$$\Delta B \le m_0 \sqrt{\frac{k\ s\ j_c}{|\P\ j_c / \P\ T|}} = B_j \qquad (3)$$

where $\sigma$ is the electrical conductivity and $\kappa/C_v$ is the thermal diffusion coefficient. In this case the typical field interval value $B_j$ is to be different and the distance in field between the successive flux jumps is not constant either.

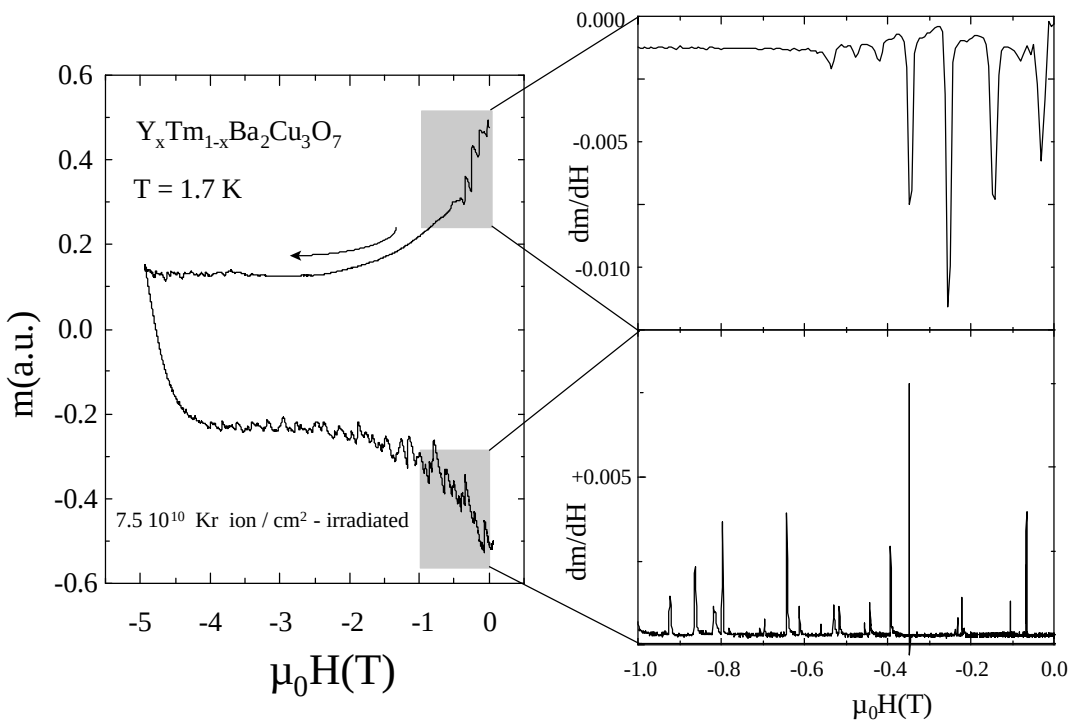

Fig. 4. The magnetization versus magnetic field for a 7.5 1010 ion/cm2 Kr-irradiated ($Y_xTm_{1-x})Ba_2Cu_30_7$ single crystal at temperature T=1.7K (the graph at the left).

We observe also that the $\chi$(H) peaks in the rising and lowering field branch are different. In the rising field branch (large dB/dt) the peaks are wider, indicating larger amounts of jumping flux.

To shed some light on the irregular behavior of flux jumps, we performed a statistical analysis of the jumps in the lowering field branch of figure 4 (lowest experimental sweep rate, T=1.7 K). Figure 5 shows this analysis. The main graph shows the jump distribution as a function of the change in magnetization. From the latter we observe that most of the jumps are rather small (incomplete) and giant flux jumps are rather sporadic. The line is a fit by an exponential decay. The insert shows the distribution of as a function of field distance between two successive jumps. Jumps with intervals of $\Delta$B~0.03 T are most common for this particular experiment.

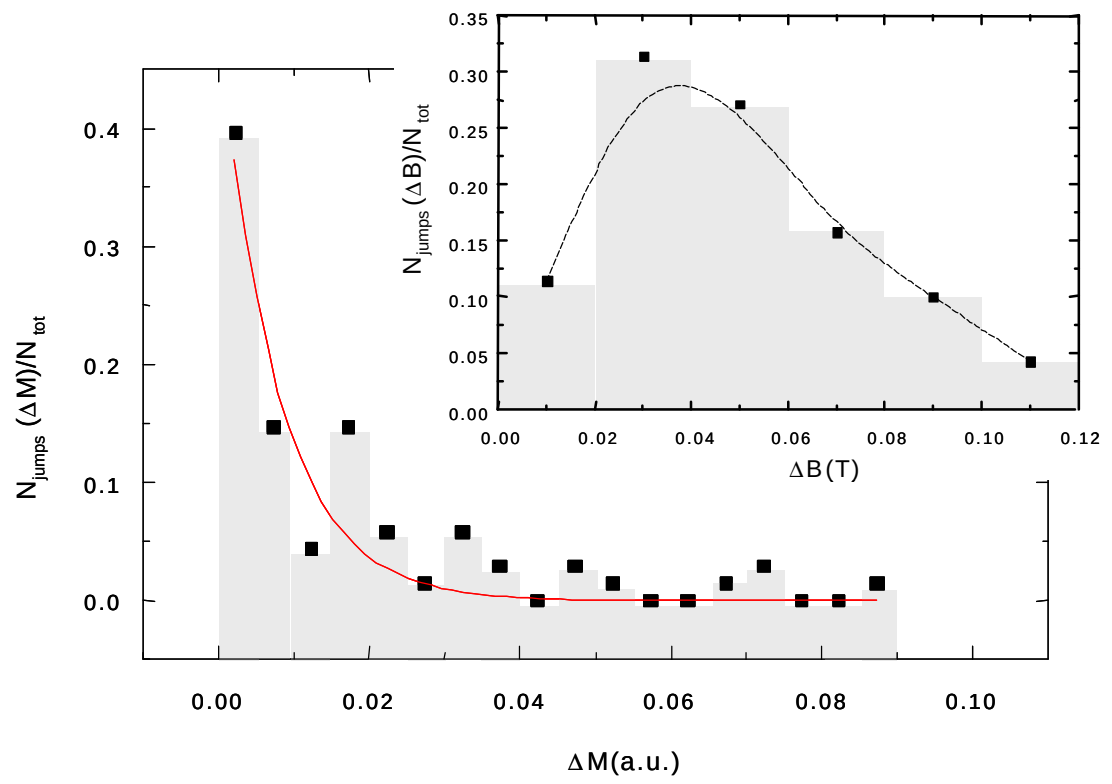

Fig. 5. Statistical analysis of the flux jumps observed in the lowering field branch in figure 4.

## Acknowledgements

This work has been supported by the Belgian IUAP, the Flemish GOA and FWO-programmes, and by the ESF programmes VORTEX. The authors wish to thank F. Nori for useful discussions.